\documentclass[fleqn,usenatbib]{mnras}

\usepackage{newtxtext,newtxmath}

\usepackage[T1]{fontenc}

\DeclareRobustCommand{\VAN}[3]{#2}
\let\VANthebibliography\thebibliography
\def\thebibliography{\DeclareRobustCommand{\VAN}[3]{##3}\VANthebibliography}

\usepackage{graphicx}	
\usepackage{amsmath}	

\title[Time delay induced by plasma in strong lens systems]{Time delay induced by plasma in strong lens systems}

\author[G. S. Bisnovatyi-Kogan and O. Yu. Tsupko]{
Gennady S. Bisnovatyi-Kogan$^{1,2}\thanks{E-mail: gkogan@iki.rssi.ru (GSBK); ORCID iD: https://orcid.org/0000-0002-2981-664X}$ and
Oleg Yu. Tsupko$^{1}$\thanks{E-mail: tsupko@iki.rssi.ru, tsupkooleg@gmail.com (OYuT); ORCID iD: https://orcid.org/0000-0002-2159-8350}
\\
$^{1}$Space Research Institute of Russian Academy of Sciences, Profsoyuznaya 84/32, Moscow 117997, Russia\\
$^{2}$National Research Nuclear University MEPhI (Moscow Engineering Physics Institute), Kashirskoe Shosse 31,\\ \qquad Moscow 115409, Russia
}

\date{Accepted XXX. Received YYY; in original form ZZZ}

\pubyear{2022}

\begin{document}
\label{firstpage}
\pagerange{\pageref{firstpage}--\pageref{lastpage}}
\maketitle

\begin{abstract}
If the gravitational lens is surrounded by non-homoheneous plasma, in addition to the vacuum gravitational deflection, chromatic refraction occurs. Also, the speed of signal propagation decreases compared to vacuum. In this article, we investigate analytically the time delay in the case of gravitational lensing in plasma, focusing on strong lens systems. We take into account the following contributions: geometric delay due to trajectory bending in the presence of both gravity and plasma; potential delay of the ray in the gravitational field of the lens; dispersion delay in the plasma due to decrease of speed of light signal in the medium. We consider singular isothermal sphere as a model of gravitational lens, and arbitrary spherically symmetric distribution of surrounding plasma. For this scenario, plasma corrections for the time delay between two images are found in compact analytical form convenient for estimates. We discuss also the possible influence of the plasma on the value of the Hubble constant, determined from observations of the time delay in strong lens systems.
\end{abstract}




\section{Introduction}

As one of its effects, gravitational lensing results in a time delay of the rays compared to unlensed propagation. In strong lens systems with multiple images, different images have different delays, and the time delay between images can be measured. Observations of the time delay make it possible to determine the Hubble constant.

If the gravitational lens is surrounded by non-homogeneous plasma, chromatic refraction occurs, in addition to the vacuum gravitational deflection.\footnote{It should be noted that even if the plasma is homogeneous, i.e., there is no refraction, the deflection of light will be different from the vacuum and chromatic. This issue has been discussed in detail in our papers \citep{BK-Tsupko-2009,BK-Tsupko-2010,Tsupko-BK-2013}, and the corresponding corrections have been calculated. A similar effect will be observed in other media with dispersion \citep{Tsupko-2021}. Usually these corrections are less than the corrections connected with refraction, and in the present paper we neglect them. For more details, see our review \citep{BKTs-Universe-2017}.} As a result, various chromatic effects can be expected (see next Section for a description of the state of research). Also, the speed of signal propagation decreases compared to vacuum. This manifests itself, for example, in the measured time delay of the signal at different radio frequencies from pulsars in the interstellar medium.

The time delay between images in presence of plasma around a gravitational lens (that is, with the simultaneous influence of gravity and the plasma medium) has not yet been widely discussed in the literature. Combined effect of plasma and weak gravitational field on propagation of signals was considered in \citet{Muhleman-1966}. They discussed the signal delay at radio frequencies in solar-corona plasma, in relation to earlier suggested Shapiro's experiment \citep{Shapiro-1964}. See also \citet{Muhleman-1970}.

Much later, already after creation of the developed gravitational lens theory, the time delay in the case of gravitational lensing in plasma has been briefly discussed by \citet{Er-Mao-2014}. They have modeled the strong lens system and, in particular, numerically estimated the time delay between images and considered its influence on measuring of the Hubble constant value. A related discussion is presented in the article of \citet{Er-2020}. The authors consider the time delay for the light propagation in the plasma, i.e. do not consider gravitational effects.\footnote{To avoid confusion, we would like to remind that three different concepts should be distinguished: (i) conventional vacuum gravitational lensing, in which the deflection angle is determined by the gravitational deflection of rays by massive objects in vacuum, for example, Einstein's formula for small angles; (ii) diverging plasma lensing, in which the deflection is caused by refraction of light rays on compact plasma inhomogeneties (no gravity effects here); (iii) gravitational lensing in presence of plasma, in which the weak or strong deflection of light due to gravity is combined with the chromatic deflection in the plasma. Effects of gravity and plasma can be completely separated only for weak deflection case. For strong gravity regime (not considered in this article), self-consistent approach should be applied, see Section \ref{sec:overview-literature} for more details.} 
They take into account both the dispersion delay and plasma geometric delay. A detailed comparison of vacuum gravitational lensing and plasma lensing (notion used for refraction on compact plasma inhomogeneities, \citet{Clegg-1998, Tuntsov-2016, Vedantham-2017, Dong-2018, Er-2020}) is presented in the article of \citet{Wagner-2020}. Just recently, the effects of plasma on the time delay for strongly lensed fast radio bursts has been investigated by \citet{Er-Mao-2022}, where both gravitational and plasma effects are taken into account; see also \citet{Kumar-Beniamini-2022}.

In this article, we investigate the time delay in the case of gravitational lensing in plasma, focusing on strong lens systems. We consider this subject completely analytically and find compact expressions for plasma corrections which are convenient for making estimates. By 'strong lens system' and 'strong lensing', we mean here a observational situation where a gravitational lens produces multiple images. To model such systems found in observations, it is always sufficient to use the deflection angle in the weak deflection approximation. Therefore, the notion 'strong lensing' should not be confused with 'strong deflection lensing' or 'strong field lensing' by compact objects where the deflection angles are large and can produce higher-order images \citep[e.g.,][]{Tsupko-BK-2013}.

In a weak deflection approximation, the total deflection angle can be calculated as the sum of vacuum gravitational deflection, and contribution of the refractive plasma deflection. In this approximation, the time delay of lensed ray (relative to the undeflected straight line ray) can be presented as a sum of the following contributions: geometric delay due to trajectory bending in the presence of both gravity and plasma; potential delay of the ray in the gravitational field of the lens; dispersion delay in the plasma due to decrease of speed of light signal in the medium; see also \citet{Er-Mao-2022}.

We calculate the time delay between two images in the following scenario: gravitational lens is given by the singular isothermal sphere (SIS) model, while the plasma component is given by arbitrary spherically symmetric distribution. Plasma corrections to the image positions and the time delay are found in a compact analytical form. We discuss also a possible influence of plasma effects on the value of the Hubble constant, determined from observations of the time delay in strong lens systems.

The paper is organized as follows. Next Section is a brief overview of the current state of research of gravitational lensing in plasma. In Section \ref{sec:time-delay-general} we present the general expression for the time delay under considered approximations. In Section~\ref{sec:sis-images}, the plasma corrections for the image positions in case of SIS lens are calculated. In Section~\ref{sec:time-delay-sis}, the plasma corrections for the time delay for SIS lens are found. In Section \ref{sec:example} we present the example of application of our formulas. Section \ref{sec:conclusions} is our Conclusions.

\section{Studies of gravitational lensing in plasma: brief review of research}
\label{sec:overview-literature}

The simplest approach for considering gravitational lensing in plasma is to use the linearized approximation, when both vacuum gravitational deflection and refraction in an inhomogeneous plasma are small, and two deflection angles are written completely separated from each other. Such method was considered in \citet{Bliokh-Minakov-1989}. The same way of deflection angle calculation was used for numerical modelling of different effects in strong lens systems by \citet{Er-Mao-2014} and for investigation of microlensing effect in plasma by \citet{Tsupko-BK-2020, Sun-2022}. The discussion of different scenarios is also presented in series of papers of \citet{BK-Tsupko-2009, BK-Tsupko-2010, BK-Tsupko-2015}. In the current paper we are also working in such approximation for calculation of the deflection angle.

If a more accurate  account of plasma effects is necessary, due to physical conditions of the problem, a self-consistent approach should be used. This means that the total deflection angle (determined by gravity and plasma) should be derived from the same theory. The propagation of rays in the curved space in  presence of a medium was considered by \citet{Synge-1960} and \citet{Perlick-2000}, see also \citet{Bicak-1975} and \citet{Kulsrud-Loeb-1992}. On the basis of Synge's equations, the deflection angle of ray in presence of both gravity and plasma has been investigated in the weak deflection approximation by \citet{BK-Tsupko-2009, BK-Tsupko-2010, BK-Tsupko-2015}. In particular, it has been shown that even in a homogeneous plasma where there is no refraction, the gravitational deflection angle differs from the vacuum one. 
Calculation of the deflection angle of light rays in plasma media up to higher order terms in expansion by plasma and gravity have been made in the works of Crisnejo and Gallo with coathours \citep{Crisnejo-2018, Crisnejo-Rogers-2019, Crisnejo-Jusufi-2019}. For a weak deflection in the Kerr metric, see \citet{Morozova-2013} and \citet{Crisnejo-Jusufi-2019}.

The exact expression for the deflection angle for a ray propagation in a gravitational field in the presence of plasma, was first found in the monograph of \citet{Perlick-2000}. The expression was derived for the motion in the equatorial plane of Kerr black hole. In the article of \citet{Tsupko-BK-2013}, the positions of higher-order images \citep{Darwin-1959, Virbhadra-2000, Bozza-2001, Perlick-review, BK-Tsupko-2008, BKTs-Universe-2017} were calculated analytically when lensed by the Schwarzschild black hole in a homogeneous plasma. An important contribution to the subject has been made in a series of articles by \citet{Rogers-2015, Rogers-2016, Rogers-2017a, Rogers-2017b}, where various effects near compact objects have been considered. Recently, spatial dispersion of light rays propagating through a plasma in Kerr space-time was investigated in \citet{Kimpson-2019a}, see also \citet{Kimpson-2019b}. The exact expression for the deflection angle in gravitational field in presence of arbitrary medium with a spherical symmetry was found by \citet{Tsupko-2021}.

Recent studies of gravitational lensing in plasmas in the strong gravity regime are also focused on the shadow \citep{Falcke-2000, Cunha-Herdeiro-2018, Perlick-Tsupko-2022} of black holes. The influence of the plasma medium on the propagation of rays and on the size and shape of the black hole shadow was calculated for the Schwarzschild black hole in \citet{Perlick-Tsupko-BK-2015}, and for the Kerr black hole in \citet{Perlick-Tsupko-2017}, see also subsequent papers \citet{Huang-2018, Yan-2019, Babar-2020, Badia-Eiroa-2021, Badia-Eiroa-2022, Chowdhuri-2021, Li-2022, Zhang-Yan-2022, Briozzo-Gallo-Madler-2022}. The light propagation in plasma (incl. separability of the Hamilton-Jacobi equation and black hole shadow) was considered in a general case of axially symmetric and stationary spacetime by \citet{Bezdekova-2022}.

Interesting results are obtained by \citet{Briozzo-Gallo-2022} who have generalized the approximate formula of \citet{Beloborodov-2002} 
for gravitational bending of light near compact objects
by including plasma corrections.
Diffraction of the light by the gravity of the Sun and the solar corona is discussed by \citet{Turyshev-2019a, Turyshev-2019b}.
For propagation of light through magnetized plasma in presence of gravity see \citet{Br-Eh-1980}, \citet{Br-Eh-1981a}, \citet{Br-Eh-1981b}, \citet{Brod-Blandford-2003a}, \citet{Brod-Blandford-2003b}. For some other recent studies see \citet{Perlick-Schwarz-2017}, \citet{Ovgun-2019}, \citet{Bratislava-2019}, \citet{Matsuno-2021}, \citet{Chainakun-2022, Guerrieri-Novello-2022}. Review of different plasma effects, incl. strong lens systems and shadow is presented by \citet{BKTs-Universe-2017}. Recently, \citet{Crisnejo-arXiv-2023, Crisnejo-Thesis, Ulla-Thesis} have considered strong lens systems given by singular isothermal ellipsoid and have studied the plasma effects analytically using perturbative approach.

Plasma lensing studies have been also carried out in \citet{Clegg-1998}, \citet{Tuntsov-2016}, \citet{Cordes-2017}, \citet{Vedantham-2017}, \citet{Dong-2018}. In a series of papers of Er and Rogers \citep{Er-Rogers-2018, Er-Rogers-2019a, Er-Rogers-2019b, Er-2020}, the formalism of gravitational lensing has been successfully applied to consider different models of plasma lenses. Interesting discussion and comparison of vacuum gravitational lensing and plasma lensing can be found in the paper of \citet{Wagner-2020}.

\section{Time delay in the case of simultaneous presence of gravity and plasma: general expression}
\label{sec:time-delay-general}

In this Section, we will consider the time delay in presence of both gravitational lens and plasma. We write the total deflection angle of light ray as the sum of vacuum gravitational deflection and refractive plasma deflection:
\begin{equation} \label{physical-angle}
\hat{\alpha} = \hat{\alpha}_{grav} + \hat{\alpha}_{refr} \, ; \quad \hat{\alpha}_{grav}, \hat{\alpha}_{refr} \ll 1 \, .
\end{equation}
Both angles are assumed to be small and independent on each other. Effects of gravity and plasma are taken into account in the linear order only, all mixed and higher-order terms are neglected here.

Working in the same order of approximation, we take into account the following contributions in time delay, as compared to straight-line propagation in vacuum in absence of gravity:

(i) the geometric delay $\Delta t^{geom}$ associated with additional path length due to the bending of the trajectory in the presence of both gravity and plasma;

(ii) the potential delay $\Delta t^{pot}$ of the ray caused by time retardation of the ray while moving in the gravitational field of the lens;

(iii) the dispersion delay $\Delta t^{disp}$ in the plasma associated with a decrease of the signal velocity in the medium.

This approach can be compared with the vacuum gravitational lensing case (when geometrical and potential delays are taken into accound) and plasma lensing case (when geometrical and dispersive delay are used).

\begin{figure}
\begin{center}
\includegraphics[width=0.45\textwidth]{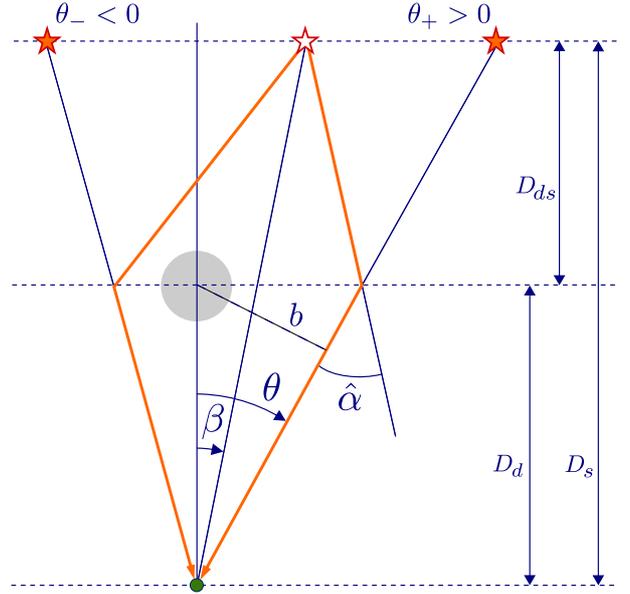}
\end{center}
\caption{(COLOR ONLINE) Standard scheme of gravitational lensing. The light ray from the distant source at the angular position $\beta$ is deflected by the lens at the 
angle $\hat{\alpha}$ and comes to the observer at the angle $\theta$. Primary image is denoted as $\theta_+$, and secondary image as $\theta_-$. $D_d$ is the distance between the lens and the observer, $D_s$ defines the distance between the distant source and the observer, and $D_{ds}$ is the distance between the source and the lens.}
\label{fig:sis-lens}
\end{figure}

We assume spherical symmetry of both the lens and the surrounding plasma. Therefore, one-dimensional variables can be used in the lens equation and the time delay expressions. Note that in general case, positions of source and images are described by two dimensional quantities in the source plane and lens plane, correspondingly. However, for axially symmetric cases, it is possible to switch to one-dimensional values in the lens equation which will be now defined in the plane where all rays forming images are located (plane of picture in 
Fig.\ref{fig:sis-lens}).

The geometrical delay is \citep[e.g.,][]{GL1, GL2, Keeton-book, Dodelson-book}:
\begin{equation} \label{geom-delay}
\Delta t^{geom}(\theta) = \frac{1+z_d}{c} \frac{D_d D_s}{D_{ds}}   \frac{( \theta - \beta )^2}{2}   \, .
\end{equation}
Here $\beta$ is the angular position of the source, $\theta$ is the position of the image, $z_d$ is the redshift of the lens, $D_i$ are angular diameter distances (Fig.\ref{fig:sis-lens}).
The form of expression (\ref{geom-delay}) is universal in the sense that it can be used for deflection caused by any physical reason. It is used not only in vacuum gravitational lensing but also in plasma lensing \citep{Er-2020, Wagner-2020}. The importance of the geometric delay in comparison with potential delay was discussed in our paper \citep{Tsupko-BK-Rogers-Er-2020}, see also \citet{Hackmann-Dhani-2019}.

The potential delay is \citep[e.g.,][]{GL1, GL2, Keeton-book, Dodelson-book}:
\begin{equation}
\Delta t^{pot}(\theta) = - \frac{1+z_d}{c} \frac{D_d D_s}{D_{ds}} \, \psi(\theta) \, + \, \mbox{const} \, .
\end{equation}
Here $\psi(\theta)$ is the deflection potential \citep{GL1,GL2} (lens potential, \citet{Keeton-book}) that depends on the mass distribution in the lens and is defined so that its gradient gives the deflection angle:
\begin{equation} \label{gradient}
\boldsymbol{\alpha}=  \boldsymbol{\nabla} \psi, \quad \mbox{where} \quad \boldsymbol{\alpha} = \frac{D_{ds}}{D_s} \boldsymbol{\hat{\alpha}} \, . 
\end{equation}

The dispersive delay is considered here for cold non-magnetized plasma with the refractive index $n$:
\begin{equation}
n^2 = 1 - \frac{\omega_p^2}{\omega^2} \, , \quad \omega_p^2 = \frac{4 \pi e^2}{m_e} N_e \, ,
\end{equation}
where $\omega_p$ is the plasma electron frequency, $\omega$ is the photon frequency (locally measured), $m_e$ and $e$ are the electron mass and charge, $N_e$ is the electron number density in plasma. Working in the linearized approximation with separated gravitational and plasma terms, we can neglect the change of photon frequency due to the gravitational field (gravitational redshift) in the terms containing plasma. Note that this cannot be neglected if it is necessary to calculate the plasma effects more precisely, for example, when finding corrections to the vacuum gravitational deflection due to the presence of homogeneous plasma \citep{BK-Tsupko-2009, BK-Tsupko-2010}, or in case of black hole shadow calculation \citep{Perlick-Tsupko-BK-2015, Perlick-Tsupko-2017, Perlick-Tsupko-2022}. In addition, if one is interested in the frequency of observation $\omega_0$, the following relation must be taken into account: $\omega=(1+z_d) \, \omega_0$, where $z_d$ is the lens redshift, see, e.g., \citet{Crisnejo-arXiv-2023, Cordes-2017}.  

With usual assumption, $\omega \gg \omega_p$, one finds:
\begin{equation}
n \simeq 1 - \frac{\omega_p^2}{2 \omega^2} = 1 - \frac{2 \pi e^2}{m_e \omega^2} N_e \, ,
\end{equation}
and for delay of ray in plasma in comparison with vacuum propagation we write:
\begin{equation}
\Delta t = \frac{1}{c} \int \left( \frac{1}{n} - 1\right) dl \simeq \frac{2 \pi e^2}{c m_e \omega^2} N_{int} \, ,
\end{equation}
where
\begin{equation}
N_{int} = \int N_e dl \, .
\end{equation}
Here $N_{int}$ is the projected electron density along the line of sight, usually referred to as the dispersion measure DM. Together with the cosmological factor we find finally:
\begin{equation}
\Delta t^{disp} =  \frac{1+z_d}{c} \frac{2 \pi e^2}{m_e \omega^2} N_{int} \, .
\end{equation}

As a result, the formula for the time delay becomes (see also \citet{Er-2020}, \citet{Er-Mao-2022}):
\begin{equation} \label{time-delay-full}
\Delta t(\theta) = \frac{D_{\Delta t}}{c}  \left[    \frac{1}{2} ( \theta - \beta )^2  - \psi(\theta)  \right]  + \frac{1+z_d}{c} \frac{K_e}{2 \omega^2} N_{int}(|\theta|) \, ,
\end{equation}
where we use a notation $K_e \equiv 4 \pi e^2/m_e$.
The variable
\begin{equation} \label{time-delay-distance}
D_{\Delta t} = (1+z_d) \frac{D_d D_s}{D_{ds}}
\end{equation}
is known as time-delay distance \citep{Suyu-2010, Suyu-2018}.

The formula (\ref{time-delay-full}) is general in the sense that any distribution of gravitating mass in the lens can be considered (which defines the function $\psi(\theta)$ together with the deflection angle, according to eq.(\ref{gradient})) and any spherically symmetric distribution of surrounding plasma given by $N_e$ and $N_{int}(|\theta|)$ can be used. We also note that the dispersive term is present both in homogeneous and not-homogeneous plasma, but it is not related to correction to gravitational deflection in homogeneous plasma discussed in the Introduction.



With the simultaneous presence of gravity and plasma, the geometric delay has a curious feature. Gravity and plasma compete with each other, acting in opposite directions. Therefore, if the angle of gravitational deflection and the angle of refraction exactly cancel each other for some image, the trajectory becomes straight and the geometric delay for this image can be equal to zero.

\section{Plasma corrections to the image positions in the case of SIS lens}
\label{sec:sis-images}

In this Section, we analytically calculate the plasma corrections to image positions in strong lens system.


The simplest model suitable for description of lensing by a galaxy or cluster is the singular isothermal sphere (SIS). For this lens model we have the following density profile $\rho_{grav}$, the deflection angle $\hat{\alpha}_{grav}$, the Einstein radius $\theta_E$ and the deflection potential $\psi(\theta)$:
\begin{equation} \label{SIS-rho}
\rho_{grav}(r) = \frac{\sigma^2}{2 \pi G r^2} \, , \quad \hat{\alpha}_{grav} = 4\pi \frac{\sigma^2}{c^2} \, ,
\end{equation}
\begin{equation} \label{SIS-param}
\theta_E = 4 \pi \left( \frac{\sigma}{c} \right)^2  \frac{D_{ds}}{D_s} \, , \quad  \psi(\theta) = \theta_E |\theta| \, .
\end{equation}
Here $\sigma$ is the velocity dispersion. For more details, see, e.g., \citet{GL1}, \citet{GL2}, \citet{Dodelson-book}, \citet{Keeton-book}. For analytical studies of plasma effects in strong lens systems with singular isothermal ellipsoid lens, we refer to \citet{Crisnejo-arXiv-2023, Crisnejo-Thesis, Ulla-Thesis}.

We assume also that the gravitating lens is surrounded by non-homogeneous spherically symmetric plasma, which leads to the refractive deflection $\hat{\alpha}_{refr}$ of light ray. Our subsequent calculations are appropriate for arbitrary law of plasma distribution.
In such an approach it becomes possible to: neglect the mass of plasma particles, or take into account the gravity created by them. In Sec.\ref{sec:example} we will consider the example when the gravitating mass is given by dark matter particles together with plasma particles, and there is the additional refractive deflection on this plasma. Alternatively, in frame of our approach, one could completely neglect the contribution of plasma particles to total gravitating mass.

Having the deflection angle $\hat{\alpha}$ in (\ref{physical-angle}) as the function of impact parameter $b$, we write $b=D_d \! \cdot \! |\theta|$ and introduce the reduced angle as
\begin{equation} \label{norm-alpha-def}
\alpha(|\theta|) = \frac{D_{ds}}{D_s} \hat{\alpha} (D_d \! \cdot \! |\theta|) =
\end{equation}
\[
= \frac{D_{ds}}{D_s} \hat{\alpha}_{grav} (D_d \! \cdot \! |\theta|) + \frac{D_{ds}}{D_s} \hat{\alpha}_{refr} (D_d \! \cdot \! |\theta|)  \, .
\]
Here the expression $D_d \! \cdot \! |\theta|$ is the argument of the functions.

With the gravitational deflection angle $\hat{\alpha}_{grav}$ from (\ref{SIS-rho}) and the Einstein angular radius $\theta_E$ from (\ref{SIS-param}), we find: 
\begin{equation} \label{norm-alpha-02}
\alpha(|\theta|) =  \theta_E  - B_{\omega}(|\theta|)  \, .
\end{equation}

Here, for convenience, we have introduced the function
\begin{equation} \label{B-omega-def}
B_\omega(|\theta|) = - \frac{D_{ds}}{D_s} \alpha_{refr}(D_d \! \cdot \! |\theta|) \, ,
\end{equation}
which is positive for density profiles falling with radius and diverging refractive deflection. By physical meaning, $B_\omega(|\theta|)$ is the refractive deflection angle normalized by the distances ratio $D_{ds}/D_s$ and taken with opposite sign. Similar notation was used in our previous paper \citep{Tsupko-BK-2020}.

To correctly describe negative $\theta$, we write the lens equation as \citep[e.g.,][]{Suyu-2012, GL2, Tsupko-BK-2020}:
\begin{equation}
\beta = \theta - \alpha(|\theta|) \frac{\theta}{|\theta|} \, .
\end{equation}

Finally, we have the lens equation with both gravitational and refractive contributions as
\begin{equation} \label{lens-eq-01}
\beta = \theta - \theta_E \frac{\theta}{|\theta|} + B_{\omega}(|\theta|) \frac{\theta}{|\theta|} \, .
\end{equation}


In order to solve the lens equation completely analytically, we will further also assume that the plasma effect is small compared to gravity:
\begin{equation}
|\hat{\alpha}_{refr}| \ll |\hat{\alpha}_{grav}| \, .
\end{equation}
This leads to condition:
\begin{equation} \label{B-omega-condition}
B_\omega(|\theta|) \ll \theta_E \, .
\end{equation}

With the condition (\ref{B-omega-condition}), the equation (\ref{lens-eq-01}) can be solved perturbatively. To do this, we introduce a bookkeeping parameter $\varepsilon$ associated with plasma terms, expand on it in a series, and at the end we will put it equal to unity. The equation (\ref{lens-eq-01}) can be written as
\begin{equation} \label{lens-eq-02}
\beta = \theta - \theta_E \frac{\theta}{|\theta|} + \varepsilon B_{\omega}(|\theta|) \frac{\theta}{|\theta|} \, ,
\end{equation}
and we look for a solution in the form of
\begin{equation} \label{look-solution}
\theta = \theta^{(0)} + \varepsilon \, \theta^{(1)} \, .
\end{equation}
The zero-order term is a vacuum solution, and the first-order term describes a linear correction due to plasma presence.

For $\theta>0$, the equation (\ref{lens-eq-02}) becomes:
\begin{equation} \label{lens-eq-03}
\beta = \theta - \theta_E  + \varepsilon B_\omega(\theta) \, .
\end{equation}
Substituting (\ref{look-solution}) into (\ref{lens-eq-03}), we find for angular position $\theta_+$ of the primary image:
\begin{equation}
\theta_+^{(0)} = \theta_E + \beta \, ,
\end{equation}
\begin{equation}
\theta_+^{(1)} = -B_+ \, , \quad \mbox{where} \; B_+ \equiv B_\omega(\theta_+^{(0)}) = B_\omega(\theta_E+\beta) \, ,
\end{equation}
\begin{equation} \label{theta-plus}
\theta_+ = \theta_E + \beta - B_+  \, .
\end{equation}

Similarly, for the secondary image ($\theta_-<0$) we find:
\begin{equation} \label{sec-image-0}
\theta_-^{(0)} = - \theta_E + \beta  \, ,
\end{equation}
\begin{equation}
\theta_-^{(1)} = B_-  \, , \quad \mbox{where} \; B_- \equiv B_\omega(|\theta_-^{(0)}|) = B_\omega(\theta_E-\beta) \, ,
\end{equation}
\begin{equation} \label{theta-minus}
\theta_- = - \theta_E + \beta + B_-  \, .
\end{equation}
Solutions of zero-order (vacuum) are known from the literature \citep{GL1, GL2, Dodelson-book, Keeton-book}. Here we find first-order plasma corrections analytically.
The secondary image exists only for $\beta<\theta_E$, as can be seen from Eq.(\ref{sec-image-0}). With $\beta \to \theta_E$, magnification factor of secondary image goes to zero, and it disappears after $\beta$ becomes bigger than $\theta_E$, see, e.g., \citet{Dodelson-book} for SIS lens in vacuum.

\section{Time delay between two images in case of SIS lens in presence of plasma}
\label{sec:time-delay-sis}

In this Section, we analytically calculate the time delay between primary and secondary images for SIS lens surrounded by plasma. With substitutions of Eq.(\ref{SIS-param}), the general expression (\ref{time-delay-full}) for the time delay becomes:
\begin{equation} \label{delay-sis}
\Delta t(\theta) = \frac{D_{\Delta t}}{c}  \left[    \frac{1}{2} ( \theta - \beta )^2  - \theta_E |\theta|  \right]  + \frac{1+z_d}{c} \frac{K_e}{2 \omega^2} N_{int}(|\theta|) \, .
\end{equation}

In order to find expressions for the time delay for a particular image, we must substitute the expressions (\ref{theta-plus}) and (\ref{theta-minus}) found earlier.
After substitution into (\ref{delay-sis}), we keep terms of the first order in plasma. By this way, we find the plasma corrections to all three terms in (\ref{delay-sis}). 
By coincidence, for this model of lens, linear plasma corrections for geometrical and potential terms of time delay exactly cancel each other. This happens because with the inclusion of the plasma, the light deflection becomes smaller and ray goes closer to the lens, so the geometric delay decreases but the potential delay increases, and for this particular distribution of matter two corrections are of the same magnitude but opposite signs. Finally, for primary image we have:
\begin{equation} 
\Delta t_+ = \frac{D_{\Delta t}}{c}  \left[   - \frac{1}{2} \theta_E^2  - \theta_E \beta  \right]  + \frac{1+z_d}{c} \frac{K_e}{2 \omega^2} N_{int}(\theta_E+\beta) \, ,
\end{equation}
whereas for secondary image:
\begin{equation} 
\Delta t_- = \frac{D_{\Delta t}}{c}  \left[   - \frac{1}{2} \theta_E^2  + \theta_E \beta  \right]  + \frac{1+z_d}{c} \frac{K_e}{2 \omega^2} N_{int}(\theta_E-\beta) \, .
\end{equation}
Time delay between images is:
\begin{equation} 
\Delta t_+ - \Delta t_-  = \frac{D_{\Delta t}}{c} (-2 \theta_E \beta) \, +
\end{equation}
\[
+ \, \frac{1+z_d}{c} \frac{K_e}{2 \omega^2} \left[ N_{int}(\theta_E+\beta) - N_{int}(\theta_E-\beta) \right] \, .
\]

Until now, we have only assumed that $\beta < \theta_E$ (in order for the SIS lens to produce a secondary image). Let us now additionally assume that $\beta \ll \theta_E$. This will allow us to significantly simplify the formulas.

First, we write approximately:
\begin{equation}
\Delta N_{int}(\theta_E+\beta) - \Delta N_{int}(\theta_E-\beta) \simeq 2 \beta \left. \frac{dN_{int}}{d\theta} \right|_{\theta=\theta_E} \, .
\end{equation}
Second, we remind that the deflection angle for the photon with unperturbed motion along $z$-axis is written as \citep{BK-Tsupko-2009, BK-Tsupko-2010, BK-Tsupko-2015}:
\begin{equation} \label{alpha-refr-b}
\hat{\alpha}_{refr}(b) =  \frac{K_e}{2\omega^2}   \int \limits_{-\infty}^\infty \frac{\partial N}{\partial b} \, dz \, .
\end{equation}
Then, for positive $\theta$, we write $b=D_d\theta$, and make the following transformation:

\begin{equation}
\frac{1+z_d}{c} \frac{K_e}{2 \omega^2} \, 2\beta \, \frac{dN_{int}}{d\theta} = \frac{1+z_d}{c} \,  2\beta \,  D_d \, \hat{\alpha}_{refr} = 
\end{equation}
\[
= \frac{1+z_d}{c} \frac{D_d D_s}{D_{ds}} \frac{D_{ds}}{D_s} \hat{\alpha}_{refr} \, 2\beta = - \frac{D_{\Delta t}}{c} \, B(|\theta|)  \,  2\beta \, .
\]

Finally, we find the compact expression for time delay between primary and secondary images of SIS lens surrounded by plasma:
\begin{equation} \label{delay-between}
\Delta t_+ - \Delta t_- = - \frac{D_{\Delta t}}{c}
\, 2\beta \left[  \theta_E + B_\omega(\theta_E)  \right] \, .
\end{equation}
Here $D_{\Delta t}$ is given by Eq.(\ref{time-delay-distance}), the variable $\theta_E$ is the vacuum Einstein radius of SIS lens (\ref{SIS-param}), and the function $B_\omega$ is defined in Eq.(\ref{B-omega-def}), where $\hat{\alpha}_{refr}$ is supposed to be known as the function of $b$.

Measuring the time delay between images allows one to determine the Hubble constant. Lens parameters from right-hand side of Eq.(\ref{delay-between}) are found from observations and modelling of the lens system. Time-delay distance (\ref{time-delay-distance}) is inversely proportional to Hubble constant \citep{Refsdal-1964, Schneider-1985, GL2, Suyu-2010, Wong-2020}:
\begin{equation}
D_{\Delta t} \propto \frac{1}{H_0} \, .
\end{equation}
As a result, the value of Hubble constant estimated from time delay measurements becomes a bit bigger if the plasma presence is taken into account. It should be noted however that results may vary, because here the simple model is considered, where plasma corrections to geometrical and potential terms are exactly cancelled. In more realistic models, the result will depend on signs and contributions of plasma corrections in all three terms of time delay.

In order to determine the Hubble constant, we need to measure the time delay $(\Delta t_+ - \Delta t_-)$ between images. All values (except $D_{\Delta t}$) in the right-hand side of the eq.(\ref{delay-between}) are found from modeling the mass distribution in the lens based on observational data. Thus, from eq.(\ref{delay-between}), the time delay distance $D_{\Delta t}$ can be calculated, and, correspondingly, we are able to find the Hubble constant $H_0$. See, e.g., \citet{Suyu-2018}.

\section{Example of calculation}
\label{sec:example}

Let us write the corresponding formulas for non-homogeneous plasma with power-law number density:
\begin{equation} \label{N-power}
N(r) = N_0 \left( \frac{r_0}{r} \right)^k ,
\end{equation}
where
\begin{equation}
N_0 = \mbox{const}, \; r_0 = \mbox{const}, \; k=\mbox{const} > 1 \, .
\end{equation}
In order to calculate the refractive deflection $\hat{\alpha}_{refr}(b)$ by formula (\ref{alpha-refr-b}), one needs to substitute $r=(b^2+z^2)^{1/2}$, take a partial derivative by $b$ and then integrate on $z$. It results in \citep{BK-Tsupko-2009, BK-Tsupko-2010, BK-Tsupko-2015, Bliokh-Minakov-1989}:
\begin{equation} \label{angle-refr-power}
\hat{\alpha}_{refr}(b) = - \frac{K_e}{\omega^2}  \frac{\sqrt{\pi} \,   \Gamma\left(\frac{k}{2} + \frac{1}{2}\right)}{\Gamma\left(\frac{k}{2}\right)} \, N_0 \left(\frac{r_0}{b}\right)^k \, ,
\end{equation}
with $\Gamma$-function
\begin{equation}
\Gamma(x) = \int \limits_0^\infty t^{x-1} e^{-t} dt \,  .
\end{equation}
Correspondingly, the function $B_\omega(|\theta|)$ which characterizes the plasma influence is:
\begin{equation} \label{B-omega-power-law}
B_\omega(|\theta|) =  \frac{D_{ds}}{D_s}  \frac{K_e}{\omega^2}  \frac{\sqrt{\pi} \,   \Gamma\left(\frac{k}{2} + \frac{1}{2}\right)}{\Gamma\left(\frac{k}{2}\right)} \, N_0 \left(\frac{r_0}
{D_d \! \cdot \! |\theta|}\right)^k \, .
\end{equation}

For example, for $k=2$ we have:
\begin{equation} \label{power-law-k-2}
\hat{\alpha}_{refr}(b) = - \frac{K_e \pi}{2 \omega^2} \, N_0 \left(\frac{r_0}{b}\right)^2 \, ,
\end{equation}
\begin{equation} \label{power-law-k-2-02}
B_\omega(|\theta|) =  \frac{D_{ds}}{D_s}  \frac{K_e \pi}{2 \omega^2} \, N_0 \frac{r_0^2}{D_d^2 |\theta|^2} \, .
\end{equation}\\

As a numerical example, let us consider SIS lens with density distribution (\ref{SIS-rho}) at redshift $z_d=0.5$ and the source at some further redshift. Corresponding value of $D_d$ is calculated with the following cosmological parameters: $H_0 = 70$ (km/s)/Mpc, $\Omega_{m0}=0.3$, $\Omega_{\Lambda 0}=0.7$. 
Distances $D_s$ and $D_{ds}$ are not specified. 
With typical value $\sigma = 200$ km/s we find:
\begin{equation}
\theta_E = 1.15'' \, \frac{D_{ds}}{D_s} \, .
\end{equation}

Let us consider the simplest case. The distribution of gravitating matter is given by (\ref{SIS-rho}), and we assume that the plasma has the same (up to constant coefficient) distribution as the rest of the matter, see \citet{BK-Tsupko-2010}:
\begin{equation} \label{plasma-example}
N(r) = \frac{\rho_{grav}(r)}{\kappa m_p} \, .
\end{equation}
Here coefficient $\kappa \simeq 6$ characterizes the contribution of plasma particles in comparison with other components of matter; $m_p$ is the proton mass. For more details, see Eq.(61) and Eq.(62) in \citet{BK-Tsupko-2010}.

Refractive deflection produced by inhomogeneous plasma with number density (\ref{plasma-example}) equals to \citep{BK-Tsupko-2010}:
\begin{equation}
\hat{\alpha}_{refr}(b) = - \frac{1}{4} \frac{K_e}{\kappa m_p} \frac{\sigma^2}{G \omega^2 b^2} \, .
\end{equation}

Correspondingly, according to Eq.(\ref{B-omega-def}), the function $B_\omega(|\theta|)$ which characterizes the plasma influence is:
\begin{equation} \label{B-omega-example}
B_\omega(|\theta|) = \frac{1}{4}  \frac{D_{ds}}{D_s D_d^2}   \frac{K_e}{\kappa m_p} \frac{\sigma^2}{G \omega^2 |\theta|^2} \, .
\end{equation}

With the frequency of the observation $\nu_0 = 327 \times 10^6$ Hz, we write $\omega_0=2 \pi \nu_0$ and $\omega=(1+z_d) \, \omega_0$, and find:
\begin{equation} \label{B-omega-example-2}
B_\omega(\theta_E) = 0.000054'' \, \frac{D_{ds}}{D_s} \, .
\end{equation}

The rays forming the images have impact parameter of the order of $b=\theta_E D_d$. At given values of other parameters, we obtain that near the lens the ray passes through the number density $N_e$ of about $0.5$ cm$^{-3}$.
For bigger plasma densities, the bigger values of plasma influence can be expected. See, e.g., \citet{Er-Mao-2014} who have considered the number densities about $10$ cm$^{-3}$.\\

\section{Conclusions}
\label{sec:conclusions}

(i) The time delay in strong lens systems surrounded by plasma environment is investigated in analytical way. We take into account three components of the time delay, as compared to straight-line propagation: the geometrical delay, the potential delay in the gravitational field, dispersion delay in the plasma. See Eq.(\ref{time-delay-full}).

(ii) Strong lens system is modelled by the singular isothermal sphere model. Plasma refractive deflection is taken into account in the form of small correction to gravitational deflection. Plasma corrections to image positions are found analytically, see Eqs. (\ref{theta-plus}), (\ref{theta-minus}). The effect is negligibly small in the optical range, but can be more significant in the radio range.

(iii) Analytical expression is derived for the time delay between two images in case of SIS lens surrounded by arbitrarily distributed spherically symmetric plasma, see Eq.(\ref{delay-between}). This allows one to estimate simply the plasma effects for the particular lens.

(iv) If the difference in image positions in different bands is observable for lens under consideration, this indicates that the plasma effects need to be taken into account in the lens modeling and further applications.

\section*{Acknowledgements}

This article is partially supported by the Russian Foundation for Basic Research, project No. 20-52-12053.

\section*{Data Availability}

Data availability is not applicable to this article as no new data were created or analysed in this study.









\bsp	
\label{lastpage}
\end{document}